\documentclass[preprint2]{aastex62}
\usepackage{color}
\usepackage{url}
\usepackage{bm}

\received{MMMM DD, YYYY}
\revised{\today}
\accepted{MMMM DD, YYYY}
\submitjournal{ApJ}
\shorttitle{Onset Mechanism of the X2.2 and X9.3 flares on AR 12673}
\shortauthors{Bamba et al.}

\begin{document}

\title{Intrusion of Magnetic Peninsula toward the Neighboring Opposite-polarity Region That Triggers the Largest Solar Flare in Solar Cycle 24}

\correspondingauthor{Yumi Bamba}
\email{y-bamba@nagoya-u.jp}

\author{Yumi Bamba}
\affiliation{Institute for Advanced Research (IAR)/Nagoya University\\Furo-cho, Chikusa-ku, Nagoya, Aichi 464-8601, Japan}
\affiliation{Institute for Space-Earth Environmental Research (ISEE)/Nagoya University\\Furo-cho, Chikusa-ku, Nagoya, Aichi 464-8601, Japan}

\author{Satoshi Inoue}
\affiliation{Institute for Space-Earth Environmental Research (ISEE)/Nagoya University\\Furo-cho, Chikusa-ku, Nagoya, Aichi 464-8601, Japan}

\author{Shinsuke Imada}
\affiliation{Institute for Space-Earth Environmental Research (ISEE)/Nagoya University\\Furo-cho, Chikusa-ku, Nagoya, Aichi 464-8601, Japan}

\begin{abstract} 

The largest X9.3 solar flare in solar cycle 24 and the preceding X2.2 flare occurred on September 6, 2017, in the solar active region NOAA 12673. 
This study aims to understand the onset mechanism of these flares via analysis of multiple observational datasets from the {\it Hinode} and {\it Solar Dynamics Observatory} and results from a non-linear force-free field extrapolation.
The most noticeable feature is the intrusion of a major negative-polarity region, appearing similar to a peninsula, oriented northwest into a neighboring opposite-polarity region.
We also observe proxies of magnetic reconnection caused by related to the intrusion of the negative peninsula: rapid changes of the magnetic field around the intruding negative peninsula; precursor brightening at the tip of the negative peninsula, including a cusp-shaped brightening that shows a transient but significant downflow ($\sim100 km/s$) at a leg of the cusp; a dark tube-like structure that appears to be a magnetic flux rope that erupted with the X9.3 flare; and coronal brightening along the dark tube-like structure that appears to represent the electric current generated under the flux rope.
Based on these observational features, we propose that (1) the intrusion of the negative peninsula was critical in promoting the push-mode magnetic reconnection that forms and grows a twisted magnetic flux rope that erupted with the X2.2 flare, (2) the continuing intrusion progressing even beyond the X2.2 flare is further promoted to disrupt the equilibrium that leads the reinforcement of the magnetic flux rope that erupted with the X9.3 flare.

\end{abstract}



\section{Introduction} \label{sec:intro}

Solar eruptions, such as solar flares \citep{carrington59} and coronal mass ejections \citep[CMEs; ][]{hansen71, tousey73, gopalswamy16}, explosively release magnetic energy that is stored in the solar corona.
These energetic phenomena are mainly driven by the topological changes of magnetic fields, such as magnetic reconnection \citep{priest02}; however, the detailed mechanisms of these phenomena are not yet fully known.
One of the key considerations regarding solar eruptions is the magnetic flux rope, a bundle of helically twisted magnetic field lines and is sometimes represented by filaments on the solar surface \citep{labrosse10, mackay10}.
Several studies have proposed the onset mechanisms of the magnetic flux rope eruptions i.e., solar flares and CMEs from both observational and numerical standpoints.
Ideal magnetohydrodynamics (MHD) instability provides a strong basis for explaining the onset of solar eruptions as flares and CMEs.
The torus instability \citep{bateman78, kliemtorok06, demoulin10, kliem14} is proposed as one of the ideal cases of MHD instability that causes solar eruptions.
The threshold is determined by the decay index, $n = -(z/B_{ex})(\partial{B_{ex}}/\partial{z})$, where $B_{ex}$ denotes the horizontal component of the external field \citep{kliemtorok06}.
Although the value of $n = 1.5$ is well known as the critical value for the onset of the torus instability, other studies have proposed that eruptions can occur even when $n$ is lesser or greater than 1.5 \citep{zuccarello15, syntelis17}.
Some observational and computational studies \citep{savcheva15, vemareddy18, roudier18} employing nonlinear force-free field (NLFFF) modeling \citep{wiegelmann_sakurai_12, inoue16, cheng17, Guo17} support the proposal regarding the torus instability.
The helical kink instability \citep{gerrard01, fangibson03, torok04, torok05} is also considered to be an ideal case of MHD instability that causes solar eruptions.
This type of instability can grow when the magnetic twist becomes strong enough, and it can grow even if a twisted flux rope does not attain the critical height of the torus instability.
Some studies show that there exists a consistency between the helical kink instability mechanism and confined flares \citep{hassanin16, liu16}.
Other non-ideal MHD models propose that certain kinds of magnetic reconnection, such as magnetic break-out \citep{Antiochos99, karpen12} and tether-cutting reconnection \citep{moore01, moore06, ishiguro17}, causes the onset of solar eruptions \citep{schmieder15}.
These magnetic reconnection models emphasize that the magnetic flux rope is formed by magnetic reconnection, while the ideal MHD models emphasize the evolution of the flux rope to an eruption by the MHD instabilities.
However, in the cases of both the ideal and non-ideal MHD models, the factor responsible for triggering the initiation of the flux rope formation and eruption is not yet fully understood.

\begin{figure}[ht!]
\plotone{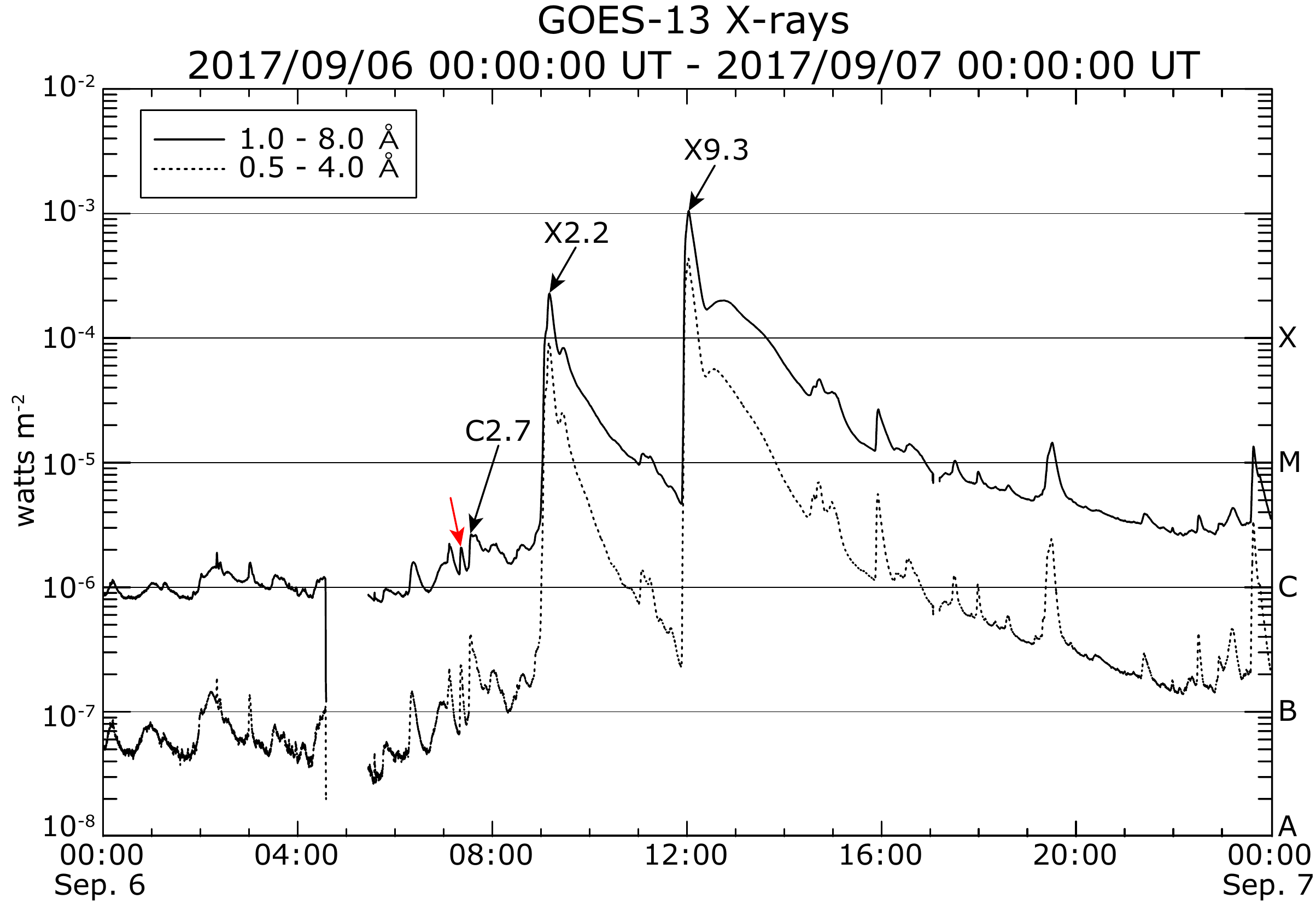}
\caption{
{\it GOES} soft X-ray light curve from 00:00 UT on September 6 to 00:00 UT on September 7, 2017.
The solid/broken lines represent 1-8 {\AA}/0.5-4 {\AA} light curves.
The onset times of the X2.2 and X9.3 flares, which are indicated by the black arrows, are 08:57 and 11:53 UT on September 6, 2017, respectively.
A small C2.7 flare occurs at 07:29 UT before the X-class flares, as indicated by the third black arrow.
The red arrow indicates soft X-ray intensity enhancement corresponding to a transient brightening and downflow observed by {\it Hinode}/EIS.
The brief data gap near 05:00 UT is due to the lunar transit.
}
\label{fig:goes}
\end{figure}

Solar active region (AR) NOAA 12673 appeared on the solar disk on August 28, 2017.
It was a small, quiet, and single sunspot containing dominant positive polarity and low negative polarity and regarded as a $\beta$-type sunspot according to the Mount Wilson (Hale) classification \citep{hale1919}.
The region did not appear to produce flares until September 2; however, sudden flux emergence was observed in the southeast and northern regions of the original bipole from September 3, 2017, and various M- and C-class flares occurred thereafter.
By September 5, 2017, this rapid flux emergence complicated the $\beta\delta\gamma$-type magnetic field structure that contains a penumbra enclosing the umbrae of both positive and negative polarities.
The $\beta\delta\gamma$-type spot is reportedly the magnetic configuration that exhibits the highest correlation with large flares, according to \citet{sammis00}; in fact, the AR 12673 produced four X-class flares on September 6, 7, and 10, 2017.
Consecutive X2.2 and X9.3 flares (Figure~\ref{fig:goes}) occurred on September 6, 2017, with rapid growth of the AR \citep{yan18, hou18}.
The X9.3 flare is the largest produced flare in solar cycle 24 and caused an increase in high-energy protons and disturbances in the Earth's ionosphere and magnetosphere.
According to \citet{redmon18}, these geo-magnetic and ionospheric disturbances disrupted HF communications in the Caribbean region during the hurricane response.
It is important to understand how and why such a large and geo-effective flare occurs, from the perspectives of both solar physics and space weather.
Therefore, in this study, we aim to understand the onset process of the geo-effective X9.3 flare and preceding X2.2 flare.

The remainder of this paper is organized as follows.
The observational data, analysis method, and NLFFF modeling are reviewed in Section~\ref{sec:analysis}.
The results of the observational data analysis are presented in Section~\ref{sec:results}.
The results are interpreted and a conceivable flare onset scenario of the X2.2 and X9.3 flares is discussed in Section~\ref{sec:discussion}.
We finally summarize and conclude the study in Section~\ref{sec:summary}.

\section{Data Analysis and Modeling} \label{sec:analysis}

\subsection{Data Description} \label{sec:analysis_data}

The X2.2 and X9.3 flares that are the focus of this study commenced at 08:57 and 11:53 UT, and peaked at 09:10 and 12:02 UT, respectively, on September 6, 2017.
{\it Hinode} \citep{kosugi07} had tracked the AR constantly from approximately 11:00 UT on September 5 until it disappeared from sight on September 14, and perfectly observed the X2.2 and X9.3 flares on September 6 using its three instruments: Solar Optical Telescope \citep[SOT, ][]{tsuneta08}, Extreme-ultraviolet Imaging Spectrometer \citep[EIS, ][]{culhane07}, and X-ray Telescope \citep[XRT, ][]{golub07}.

{\it Hinode}/SOT determined the full polarization states (Stokes-I, Q, U, and V) at 6301.5 {\AA} and 6302.5 {\AA} (\ion{Fe}{1} line) with a sampling of 21.5 m{\AA} using a Spectro-Polarimeter \citep[SP, ][]{lites13}.
We used an SP map, which was scanned from 06:14 UT on September 6, 2017, to analyze three components of the photospheric magnetic field prior to the onset of the two X-class flares.
The SP map was obtained with the fast-mapping mode observation that covered a field-of-view of 164$^{\prime\prime} \times$ 164$^{\prime\prime}$.
The sampling was 0.3$^{\prime\prime}$ along both the slit and slit-scan.

{\it Hinode}/EIS successfully observed both the X2.2 and X9.3 flares.
We investigated the local Doppler velocities of plasma related to the precursor brightening using the data from EIS.
The raster scan was repeated from 06:20 to 15:35 UT on September 6, 2017, for 10 different wavelengths.
In this study, we mainly used \ion{Fe}{14} lines corresponding to 264.79 {\AA} and 274.20 {\AA}, \ion{Fe}{16}/\ion{Fe}{23} lines corresponding to 263.3 {\AA}, and \ion{He}{2} lines corresponding to 256.3 {\AA}.
The Fe and \ion{He}{2} lines are sensitive to emissions from $log(T)\sim6.3-6.4$ plasmas and $log(T)\sim4.9$ plasmas, respectively \citep{culhane07}.
The $2^{\prime\prime} \times 120^{\prime\prime}$ slit scanned 30 positions with sparse $4^{\prime\prime}$ steps, where $4^{\prime\prime}$ is the distance between the center of the slits; hence, the scanning field-of-view covered $120^{\prime\prime} \times 120^{\prime\prime}$.
The exposure time at each position was 4s, and the spectral resolution was 22 m{\AA}.

{\it Hinode}/XRT continuously observed the AR using a standard AR observation program; this program captures coronal images in multiple wavelengths with high time cadence.
We investigated the high-temperature coronal loops that appeared to relate to the magnetic reconnection among sheared magnetic fields using the XRT images that were captures with a thin-Be/Al-poly filter covering a $384^{\prime\prime} \times 384^{\prime\prime}$ field-of-view.
The spatial resolution was $2^{\prime\prime}$, and the data cadence of each thin-Be/Al-poly images was approximately 90s.

The {\it Solar Dynamics Observatory} \citep[{\it SDO}, ][]{pesnell12} observed the X2.2 and X9.3 flares using the Helioseismic and Magnetic Imager \citep[HMI, ][]{schou12} and the Atmospheric Imaging Assembly \citep[AIA, ][]{lemen12}.
We used the {\it SDO} data to investigate the temporal evolution of the photospheric magnetic fields and the precursor activities in the solar atmosphere.
Moreover, we primarily used vector magnetic field data referred to as the Space-weather HMI AR Patch \citep[SHARP\footnote{``{\tt hmi.sharp\_720s}'' series downloaded from Joint Science Operations Center (JSOC) at Stanford University (\url{http://jsoc.stanford.edu/})}, ][]{bobra14} data series.
The SHARP data series has been calibrated assuming the Milne-Eddington atmosphere; its $180^{\circ}$ ambiguity in the horizontal component was resolved via the minimum energy method \citep{metcalf94, leka09b, leka12}.
The SHARP data was continually obtained every 12 min, and we considered the data obtained from 00:00 UT on September 6, to 00:00 UT on September 7, 2017, for use in this study.
We also used AIA level 1 data\footnote{``{\tt aia.lev1\_euv\_12s}'' and ``{\tt aia.lev1\_uv\_24s}''  series} that was obtained in the time window of 00:00 UT on September 6, to 00:00 UT on September 7, 2017.
AIA observed the AR with 10 wavelengths, but we mainly analyzed the images captured corresponding to 1600{\AA} and 171{\AA}.
We used AIA 211{\AA} and 335 {\AA} data to spatially co-align the HMI/SHARP magnetic field data and {\it Hinode}/EIS or XRT data (see details in Section~\ref{sec:analysis_hinode}).
The data cadence was 24s for 1600{\AA} and 12s for 171{\AA}, 211{\AA}, and 335{\AA}.
In addition, we used HMI level 1.5 line-of-sight magnetic field data\footnote{``{\tt hmi.M\_45s}'' series} that was observed via HMI filtergram observation for the \ion{Fe}{1} line (6173{\AA}) with a 45s cadence in the same time window as that of AIA data to superpose SHARP vector magnetograms and AIA or {\it Hinode}/EIS or XRT images.
The spatial resolution was $1.0^{\prime\prime}$ for SHARP and HMI/line-of-sight magnetograms, and $1.5^{\prime\prime}$ for AIA images.
Note that we avoided using the {\it SDO} data obtained from 06:22:34 UT to 07:53:11 UT on September 6, 2017, because these data were not of good quality owing to the lunar transit.

\subsection{Analysis Methods for Hinode Data} \label{sec:analysis_hinode}

The {\it Hinode}/SOT-SP data was calibrated via the {\tt sp\_prep} procedure \citep{lites_ichimoto13} using the Solar-SoftWare (SSW) package \citep{freeland98}.
Subsequently, vector magnetic fields were derived from the calibrated Stokes profiles based on the assumption of the Milne-Eddington atmosphere using the inversion code {\tt MERLIN} at HAO/CSAC (DOI: 10.5065/D6JH3J8D).
The minimum energy code {\tt ME0} \citep{leka09a, leka09b, leka12}, which is based on the minimum energy method of \citet{metcalf94}, was adopted to resolve the 180$^\circ$ ambiguity in the tangential component of the magnetic fields.
As highlighted by \citet{leka17, bamba18}, the three components of the photospheric magnetic field are influenced by a projection-effect; that is, the line-of-sight component of the magnetic field can be treated as the radial component only when the observing angle $\theta = 0$.
The AR 12673 was located on September 6, 2017, at S09W42, which is close to the south-east limb.
Therefore, to investigate locations and shapes of magnetic polarity inversion lines, we transformed the vector magnetic field data obtained via SOT-SP, which is in image plane on the Charge Coupled Device (CCD), to its corresponding heliocentric coordinates.
To transform the magnetic field vectors, we used the transformation matrix presented by equation (1) of \citet{gary_hagyard90}.
In this manner, we obtained the radial component ($B_z$) and the horizontal components ($B_x, B_y$) of the photospheric magnetic field vectors.

{\it Hinode}/EIS raster scan data was calibrated via the {\tt eis\_prep} procedure using the SSW package.
For reasons concerning thermal factors, there exists an orbital variation of the line position causing an artificial Doppler shift of $\pm 20 km/s$, which follows a sinusoidal behavior.
This orbital variation and wavelength calibration of the line position were corrected using the house keeping data \citep{kamio10}.
Subsequently, we first performed wavelength calibration using the \ion{Fe}{14} (264.787{\AA}) line, following which we finalized the central value $\lambda^{\prime}_{obs}$ for the observed \ion{Fe}{14} line profiles via the single Gaussian fitting.
Herein, we integrated the observed line profile over the region that was relatively quiet as per the EIS scanning field-of-view.
We then applied the single Gaussian fitting to the integrated line profile and repeated the process for the data that were scanned from 06:21:03\footnote[4]{} UT to 06:58:03\footnote{It means scan start time.} UT on September 6, 2017.
We averaged the value of fitted line center over the abovementioned time period and obtained the observed line center $\lambda_{obs\_FeXIV}$.
Subsequently, we determined {\it dw} by subtracting $\lambda_{obs\_FeXIV}$ from $\lambda_{lit\_FeXIV}$, which is the literature wavelength of the \ion{Fe}{14} line as per literature \citep{brown08}: $dw = \lambda_{lit\_FeXIV} -  \lambda_{obs}$.
Using  {\it dw}, we calibrated the other spectral lines as per $\lambda_{0} = \lambda_{lit} - {\it dw}$, where $\lambda_{lit}$ is the wavelength of the other lines based on literature \citep{brown08}, and $\lambda_{0}$ is the reference wavelength of those lines as per the EIS raster scan observation.
Following this, we identified the center of observed spectral lines $\lambda_{obs}$ via the {\tt eis\_auto\_fit} procedure using the SSW package, which applies single Gaussian fitting to the line profiles.
We then generated intensity, Doppler velocity, and line width maps for each of the observed lines.
The Doppler velocity was calculated as per $v = {\Delta\lambda}c/{\lambda_{0}}$, where $\Delta\lambda = \lambda_{0} - \lambda_{obs}$, and $c$ is the speed of light.
Using the intensity, Doppler velocity, and line width maps, we checked the locations and timings of precursor brightening occurrences relative to the two X-class flares.
We further applied double Gaussian fitting to the averaged line profile of the \ion{Fe}{14} lines (264.7{\AA} and 274.2{\AA}) at the single-slit position of the precursor brightening and measured the Doppler velocity of a local and transient flow, which is described in Section~\ref{sec:results_local}.
Note that the blend from the other lines is negligibly small for \ion{Fe}{14} (264.7{\AA}) and \ion{Fe}{16} (262.9{\AA}) \citep{young07}.

The {\it Hinode}/XRT data was calibrated via the {\tt xrt\_prep} procedure \citep{kobelski14} using the SSW package.
Subsequently, we first spatially co-aligned the XRT Al-poly images and {\it SDO}/AIA 335{\AA}, which show structures similar to each other \citep{yoshimura15}.
The AIA 335{\AA} images have been spatially well aligned with the HMI/line-of-sight images via the {\tt aia\_prep} procedure, as we described in Section~\ref{sec:analysis_SDO}.
Thus, we aligned the radial magnetic field ($B_z$) images of {\it Hinode}/SOT-SP or {\it SDO}/HMI with the line-of-sight magnetic field images.
In this manner, we finally overplotted the XRT intensities onto the $B_z$ field images or vice versa (Figure~\ref{fig:XRT_MAG}).

\subsection{Analysis Methods for SDO Data} \label{sec:analysis_SDO}

HMI/line-of-sight magnetograms and AIA data were calibrated via the {\tt aia\_prep} procedure using SSW, whereby can be rotated such that the solar EW and NS axes are aligned with the horizontal and vertical axes of the image.
The pixel scales, which are different between HMI and AIA, were resampled to the same size.
Thus, the positions of the line-of-sight magnetograms and AIA images positions were aligned.
Next, we transformed SHARP vector magnetograms to the heliocentric coordinates in the same manner for {\it Hinode}/SOT-SP.
Therefore, we obtained the radial component ($B_z$) and horizontal components ($B_x, B_y$) of the photospheric magnetic field vectors.
We then extracted an line-of-sight magnetogram, SHARP vector magnetogram, and AIA image, which were the closest in terms of time, and superposed the SHARP vector magnetic field and AIA image via the line-of-sight magnetogram.
Moreover, we superposed the magnetic polarity inversion lines of the radial component ($B_z$) of the SHARP data onto the {\it Hinode}/EIS maps corresponding to the intensity, Doppler velocity, and line width maps.
For this superposition, the {\it Hinode}/EIS maps corresponding to the \ion{Fe}{14} (264.787{\AA}) line and AIA 211{\AA} images were superposed.
The AIA 211{\AA} images and the SHARP magnetic field data had already been co-aligned via the line-of-sight magnetic field data; thus the magnetic polarity inversion lines of $B_z$ from the SHARP data were overplotted onto the {\it Hinode}/EIS maps.
We showed the magnetic polarity inversion lines on the EIS maps, which were observed during the precursor brightening (Figure~\ref{fig:EIS_SHARP}).
Then magnetic polarity inversion lines of $B_z$ fields from the SHARP were also overplotted on the {\it Hinode}/XRT images (Figure~\ref{fig:XRT_MAG}) in the same manner, although herein we employed AIA 335{\AA} instead of 211{\AA}, as we mentioned in Section~\ref{sec:analysis_hinode}.

\subsection{Nonlinear Force-free Field Modeling} \label{sec:analysis_NLFFF}

We extrapolated three-dimensional (3D) coronal magnetic fields based on the photospheric magnetic fields before the two X-class flares.
We used the modeling of the 3D coronal magnetic field using an NLFFF approximation, as described in \citet{wiegelmann_sakurai_12}, and employed the MHD relaxation method developed in \citet{inoue14, inoue16}.
The SHARP vector magnetic fields ($B_x, B_y, B_z$) observed at 08:36 and 11:00 UT on September 6, 2017, were set as the bottom boundary condition.
We solved the MHD equations in 3D space starting with a potential field, which was extrapolated from the radial component $B_z$ of the photospheric magnetic field using the Green function method \citep{sakurai82}.
Note that pressure and gravity were neglected here because the low $\beta$ approximation was well satisfied in the solar corona \citep{gary01}.
Therefore, we solved the MHD equations simplified in the zero $\beta$ approximation. 
The basic equations are as follows:
   \begin{equation}
   \rho = |{\bf B}|,
   \end{equation}
    \begin{equation}
    \frac{\partial {\bf v}}{\partial t} 
                         = - ({\bf v}\cdot{\bf \nabla}){\bf v}
                           + \frac{1}{\rho} {\bf J}\times{\bf B}
                           + \nu{\bf \nabla}^{2}{\bf v},
   \label{eq_of_mo}    
   \end{equation}
  \begin{equation}
  \frac{\partial {\bf B}}{\partial t} 
                        =  {\bf \nabla}\times({\bf v}\times{\bf B}
                        -  \eta{\bf J})
                        -  {\bf \nabla}\phi, 
  \label{in_eq}
  \end{equation}
  \begin{equation}
  {\bf J} = {\bf \nabla}\times{\bf B},
  \label{Am_low}
  \end{equation}
  \begin{equation}
  \frac{\partial \phi}{\partial t} + c^2_{h}{\bf \nabla}\cdot{\bf B} 
    = -\frac{c^2_{h}}{c^2_{p}}\phi,
  \label{div_eq}
  \end{equation}
where $\rho$, ${\bf B}$, ${\bf v}$, ${\bf J}$, and $\phi$ are plasma density, magnetic flux density, velocity, electric current density, and a convenient potential to remove errors derived from ${\bf \nabla}\cdot {\bf B}$ \citep{dedner02}, respectively.
In these equations, the length, magnetic field, density, velocity, time, and electric current density are normalized by   
$L^{*} = 244.8 Mm$,  
 $B^{*} = 2,500 G$, 
 $\rho^{*}$ = $|B^{*}|$,
 $V_{\rm A}^{*}\equiv B^{*}/(\mu_{0}\rho^{*})^{1/2}$,    
 where $\mu_0$ is the magnetic permeability,
 $\tau_{\rm A}^{*}\equiv L^{*}/V_{\rm A}^{*}$, and 
 $J^{*}=B^{*}/\mu_{0} L^{*}$,  
 respectively.
 
The pseudo density is assumed to be proportional to $|{\bf B}|$ to ease the relaxation by equalizing the Alfv$\acute{e}$n speed in space.
$\nu$ is the viscosity fixed at $1.0 \times 10^{-3}$, and the coefficients $c_h^2$, $c_p^2$ in Equation (5) are also fixed with the constant values, 0.04 and 0.1, respectively.
The initial density condition is given by $\rho = |{\bf B}|$, and the velocity is set to zero in the entire space.
The resistivity is given by $ \eta = \eta_0 + \eta_1 |{\bf J}\times{\bf B}||{\bf v}|/|{\bf B}|$ where $\eta_0 = 5.0\times 10^{-5}$ and $\eta_1=1.0\times 10^{-3}$ in non-dimensional units.
The second term is introduced to accelerate the relaxation to the force free field, particularly in weak field regions.
At the boundaries, all of velocities are fixed to zero, and the magnetic fields are fixed to the potential fields except at the bottom boundary.
More details are presented in \cite{inoue18}.

\section{Results} \label{sec:results}

\subsection{Evolution Steps of the X2.2 and X9.3 Flares} \label{sec:results_evolution}

The X2.2 and X9.3 flares occurred in an interval of approximately 3h as shown in {\it GOES}\footnote{\it{Geostationary Operational Environmental Satellite}: We employed GOES-13 because GOES-15, which was the latest GOES series as of September 2017, had brief outage of soft X-ray due to lunar transit near the onset time of the X2.2 flare.} soft X-ray light curves of Figure~\ref{fig:goes}.
The onset time was 08:57 UT for the X2.2 flare and 11:53 UT for the X9.3 flare, on September 6, 2017.
Another C2.7 flare occurred before the two X-class flares, and its onset time was 07:29 UT, on September 6, 2017.
Figure~\ref{fig:AR_evolution}(a-c) shows the temporal evolution of the radial magnetic field $B_z$ from {\it SDO}/HMI SHARP.
Panels (a), (b), and (c) are snapshots at 04:58:42 UT (approximately 4h before the X2.2 flare), 08:46:42 UT (approximately 10min before the X2.2 flare) and 11:22:42 UT (approximately 30min before the X9.3 flare).
We labeled the major polarities as N1 and N2 for the negative region, and P1 and P2 for the positive region, as shown in panel (a).
The most remarkable magnetic field change was the intrusion of a part of the N1, which is indicated by the red circle, toward the northwest direction as indicated by the red arrow in panel (a).
The region was shaped like a peninsula and intruded into the neighboring positive polarity region P1.
Henceforth, this region is called "intruding negative peninsula."
It was originally shaped as shown in panel (a), and it seemed to start the intrusion between 06:50 and 07:50 UT, when the HMI observation was not available, or the data were not of good quality because of the lunar transit.
Then, the intrusion became rapid from around 08:00 UT, i.e., from approximately 1h before the X2.2 flare onset.
The intrusion continued throughout the two X-flares (panel (b)).
Then, the intruding negative peninsula finally merged with the N2, dividing the P1, as shown in panel (c).

Precursor brightening\footnote{We defined precursor brightening as a small brightening intermittently observed at the same site on a magnetic polarity inversion lines where initial flare ribbons appeared on both sides. A physical interpretation of precursor brightening in terms of flare trigger process is discussed in \citet{bamba18}.} and initial flare ribbons of the two X-flares are observed near the intruding negative peninsula.
Panels (d-i) are snapshots of the last precursor brightening (panels (d) and (g)), initial flare ribbons (panels (e) and (h)), and enhanced flare ribbons (panels (f) and (i)) for the X2.2 and X9.3 flares, observed by {\it SDO}/AIA 1600{\AA}.
Several noticeable features were obtained by comparing the X2.2 and X9.3 flares.
First, the last precursor brightening was observed at the tip of the intruding negative peninsula before the X2.2 flare, while the location of the last precursor for the X9.3 flare was slightly south from the tip of the intruding negative peninsula, as indicated by the red arrows in panels (d) and (g).
In addition, the shape of the last precursor brightening was dot-like for the X2.2 flare, while it was elongated along the magnetic polarity inversion line between P1 and N1 for the X9.3 flare.
Second, the initial flare ribbons of the X2.2 flare appeared near the site where the last precursor brightening was observed.
The ribbons over the P1 were labeled as R1 and R2, and those over the N1 and N2 were labeled as R3, R4, and R5, as indicated by the white arrows in panel (e).
The ribbons R1-R3 were locating around the the tip of the intruding negative peninsula, although the R4 and R5 were at a slight distance from it.
The initial flare ribbons of the X9.3 flare also appeared on both sides of the location of the its precursor brightening, although the ribbons were saturated immediately, as seen in panel (h).
Third, the enhanced flare ribbons were extended further  south for the X9.3 flare, compared with those of the X2.2 flare.
Both the X-flares showed north-south (along the magnetic polarity inversion line between P1 and N1) and east-west (along the magnetic polarity inversion line between P1-N2) ribbons, as shown in panels (f) and (i).
In the X2.2 flare, the east-west ribbons R4 and R5 appeared almost simultaneously with the north-south ribbons R1-R3; R5 was fainter than the others.
Even after the ribbons were enhanced, their shape and location were not very different in panels (e) and (f).
In contrast, the X9.3 flare ribbons were extended more toward west and south directions in panel (i) than those in panel (h).
These temporal variations in precursor brightening and flare ribbons can be clearly seen in the supplemental animation of the {\it SDO}/AIA 1600 {\AA}.

\subsection{Local Structures in the X2.2 and X9.3 Flaring Site} \label{sec:results_local}

The intruding negative peninsula is likely to related to the onset of the two X-class flares, as described in Section~\ref{sec:results_evolution}.
Therefore, we further investigated the magnetic field structure and corresponding plasma dynamics near the intruding negative peninsula.
Figure~\ref{fig:SP_angles} (a) shows the three components of magnetic fields in the AR observed by {\it Hinode}/SOT-SP, approximately 3h before the X2.2 flare.
{\it Hinode}/SOT-SP successfully scanned the central part of the AR including the intruding negative peninsula, although it missed the southern half of P2.
The magnetic polarity inversion line between P1 and N1, including the intruding negative peninsula, had a strong magnetic gradient in the radial component ${B_z}$ and strong magnetic shear in the horizontal components $(B_x, B_y)$, as shown in panel (a).
The magnetic shear distribution is more clearly shown in panel (b), where a strong magnetic shear ($<-70^{\circ}$ of shear angles) is distributed along the magnetic polarity inversion line between P1 and N1.
In particular, the tip of the intruding negative peninsula was close to $-90^{\circ}$ magnetic shear as the vectors of the horizontal fields were almost parallel to the magnetic polarity inversion line in panel (a).

Interestingly, a cusp-shaped brightening was observed at the tip of the intruding negative peninsula approximately 1.5h before the X2.2 flare onset.
It was observed as a transient enhancement of soft X-ray as indicated by the red arrow in Figure~\ref{fig:goes}.
Figure~\ref{fig:EIS_SHARP} shows quicklook maps of intensity, Doppler velocity, and line width from {\it Hinode}/EIS spectroscopic observations of \ion{Fe}{14} (264.7{\AA}) line, which is sensitive to emissions from plasma at a temperature of $log(T)\sim6.3$ with overlying magnetic polarity inversion lines of the radial magnetic field $B_z$ from {\it SDO}/HMI SHARP.
{\it Hinode}/EIS field-of-view successfully covered both the north-south and east-west magnetic polarity inversion lines, and the cusp-shaped brightening can be seen in the intensity map of panel (b).
The cusp-shaped brightening bridged over the north-south magnetic polarity inversion lines between P1 and N1 at the tip of the intruding negative peninsula.
{\it Hinode}/EIS scanned the AR from west to east (from right to left in the maps) in the field-of-view for 3min, and the time in the Figure~\ref{fig:EIS_SHARP} indicates the scan start time.
Therefore, the cusp-shaped brightening was observed approximately at 07:19 UT.
The brightening was simultaneously observed by {\it SDO}/AIA in 171{\AA}, which is sensitive to emissions from plasma at a temperature of $log(T)\sim5.8$ \citep{lemen12}, with considerably higher 12min temporal resolution, as shown in panels (e-i)\footnote{Note that data of the AIA 171 {\AA} on panels (e-i) are not good quality because of lunar transit. We could still clearly identify the cusp-shaped brightening and its evolution as shown in panels (e-i).}.
The brightening started in a small region indicated by the white arrow in panel (e), and propagated to both sides; then, the cusp-shaped brightening appeared (see panels (e-i)).
This suggests that there was an energy input at the top of the cusp, and plasma fell down along the magnetic field lines.

Moreover, transient but significant downflow was observed to the west of the intruding negative peninsula, as shown in panel (c).
Because {\it Hinode}/EIS performed observations with a $2^{\prime\prime}$ slit, the red-shifted Doppler signal was only seen in one slit position.
It can be identified that the red-shifted signal corresponds to the west (right in the image, roots in P1) foot point of the cusp-shaped brightening.
The Doppler velocity was more than $80 km/s$ in the \ion{Fe}{14} line.
Significant line broadening was also observed for the cusp-shape in panel (d).
One of the reasons of line broadening is the presence of bi-directional plasma jets \citep[e.g.,][]{innes97} or turbulent motion \citep[e.g.,][]{imada08} caused by magnetic reconnection.
There has been considerable discussion on line broadening associated with magnetic reconnection.
The shape of the observed spectral line in our study is similar to the spectral line associated with magnetic reconnection discussed in a previous study (see Figure1 in \citet{innes97}).
Furthermore, coronal dimming and strong blue-shifted signals were observed in the northeast of  the field-of-view in Figure~\ref{fig:EIS_SHARP}(b-d).
At 06:21 UT, this coronal dimming had already started and was further enhanced after the X2.2 and X9.3 flares.
It is inferred that the coronal dimming related to the expanding coronal loops and corresponding upflows were observed a few days or a day before the flare onset, as suggested by \citet{Imada14}.

We analyzed the spectral line profiles at the site of the cusp-shaped brightening via double Gaussian fitting.
Figure~\ref{fig:EISprofile} shows the spectral line profiles and spectral images of \ion{Fe}{14} (264.7 {\AA} and 274.2 {\AA}), \ion{Fe}{16}/\ion{Fe}{23} (263.3 {\AA}), and \ion{He}{2} (256.3 {\AA}) lines at the single site of the red-shifted signal, shown in Figure~\ref{fig:EIS_SHARP}(c).
The vertical axis of the spectral images (a-2, b-2, c-2, and d-2) corresponding to the Y-axis of Figure~\ref{fig:EIS_SHARP}(c), and the white horizontal line-cut denotes the Y-coordinate of the strongest red-shifted signal.
The intensity profiles along the white line-cut are plotted in the left for each spectral image (a-1, b-1, c-1, and d-1).
The horizontal axis for both intensity profiles and spectral images are converted to denote velocity unit by single (in panels (c) and (d)) or double (in panels (a) and (b)) Gaussian fitting.
Evidently, both \ion{Fe}{14} lines have significant red-shifted components, as shown in panels (a-1) and (b-1).
The Doppler velocity of the second components (red-colored profiles) relative to the first components (blue-colored profiles) was approximately $103 \pm 8 km/s$ for 264.7 {\AA} and $115 \pm 16 km/s$ for 274.2 {\AA}, respectively.
Note that the errors are defined as the summation of errors due to photon noise for each component of the double Double Gaussian fitting.
\ion{Fe}{16}/\ion{Fe}{23} (263.3 {\AA}) and \ion{He}{2} (256.3 {\AA}) lines in panels (c-1) and (d-1) also show red-shifted profiles although it was not well fitted via double Gaussian fitting.
These profiles \ion{Fe}{16}/\ion{Fe}{23} and \ion{He}{2} lines were also seen to have red-shifted components with a velocity of approximately $100 km/s$.

\subsection{Global Structures in AR 12673} \label{sec:results_global}

The global coronal structure of the AR 12673 was observed using {\it Hinode}/XRT and {\it SDO}/AIA.
Figure~\ref{fig:XRT_MAG} shows the temporal variation of high-temperature coronal loops observed using {\it Hinode}/XRT, in which the magnetic polarity inversion lines of the radial magnetic fields $B_z$ from {\it SDO}/HMI SHARP or {\it Hinode}/SOT-SP are overplotted.
The red global coronal loops connect the major bipoles P1-N1, P1-N2, and P2-N1 in panel (a).
In particular, P1-N1 loops were highly sheared and almost parallel to the magnetic polarity inversion lines.
The small loop that is indicated by the white arrow in panel (a) existed in the north of the intruding negative peninsula at least from 00:00 UT on September 6, 2017.
A cusp-shaped brightening then appeared just before the C2.7 flare as seen in panel (b).
This cusp-shaped brightening \citep[c.f. ][]{tsuneta92} corresponded to the one that was observed by {\it Hinode}/EIS and {\it SDO}/AIA 171 {\AA} in Figures~\ref{fig:EIS_SHARP} and \ref{fig:EISprofile}, and the transient enhancement of soft X-ray as denoted by the red arrow in Figure~\ref{fig:goes}.
The cusp-shaped brightening remained until the X2.2 flare onset, following which, a dark tube-like feature was observed just before the X9.3 flare onset, as denoted by the sky-ble arrows in panels (c) and (d).
This tube-like dark feature was elongated north-south and was along the magnetic polarity inversion lines between P1 and N1.

Figure~\ref{fig:AIA171fluxrope} shows the coronal loops of lower temperature than those shown in Figure~\ref{fig:XRT_MAG}, observed by {\it SDO}/AIA 171 {\AA}.
Panels (a) and (b) are snapshots of HMI radial magnetic field and AIA 171 {\AA} image, respectively, taken approximately 45min before the X2.2 flare onset.
There was a small loop that is bridging over the magnetic polarity inversion line between P1 and N1 at the tip of the intruding negative peninsula, as indicated by the red arrow in panel (b).
This small loop corresponded to the cusp-shaped brightening seen in Figures~\ref{fig:EIS_SHARP}(e-i) and \ref{fig:XRT_MAG}(b).
Moreover, another bright loop L1 is denoted by the white arrow in panel (b).
The loop L1 remained until the X2.2 flare onset, as seen in panel (c).
The initial flare ribbons, shown in Figure~\ref{fig:AR_evolution}(e), are also seen in panel (c), although the loop L1 only can be seen in the {\it SDO}/AIA 171 {\AA} image.
The loop L1 seemed to correspond the coronal loop that was rooted in P1 and N1.
The loop L2, which has the same connectivity as L1, appeared after the X2.2 flare, as denoted by the white arrow in panel (d).
The intensity of L2 gradually decayed, meanwhile a long north-south loop that is connecting P2-N2, denoted by a white arrow in panel (e), gradually appeared.
The long north-south loop expanded as is outlined by white dashed line in panel (f) and erupted simultaneously with intense emission of the X9.3 flare onset.

Interestingly, the dark tube-like structure that is shown in Figure~\ref{fig:XRT_MAG} (c) and (d) is also observed in the AIA 171 {\AA} image, as denoted by the sky-ble arrow in Figure~\ref{fig:AIA171fluxrope} (e).
This tube-like structure is along the magnetic polarity inversion lines between P1 and N1 and is also observed under the bright loop L2 in panel (d).
It was not observed in AIA 171 {\AA} image before the X2.2 flare; however, it started to be seen as gradually rise from around 10:45 UT, when the brightening of the X2.2 flare ribbons was decaying as shown in the supplemental animation.
The dark tube-like structure then became invisible because of intense emission from the X9.3 flare ribbons in panel (f).
It seemed to be erupted by the X9.3 flare because there was no clear structure such as dark tube after the X9.3 flare emission started decaying.

\section{Discussion} \label{sec:discussion}

\subsection{Conceivable Scenario of sequential X2.2 and X9.3 Flares} \label{sec:rdiscussion_scenario}

The X-class flares occurred while the intruding negative peninsula was moving toward northwest as we described in Section~\ref{sec:results_evolution}.
Our results suggest that the intrusion possibly causes variations in magnetic topologies owing to magnetic reconnection along the magnetic polarity inversion line between P1 and N1.
Figure~\ref{fig:NLFFF}(a) shows the magnetic field lines that produce the X2.2 flare as blue and magenta tubes on the radial magnetic field image.
The positive/negative foot point of magenta (blue) tubes is labeled as F1/F2 (F3/F4), as shown in panel (b), in which the back ground is the image of the initial flare ribbons in AIA 1600 {\AA}.
The negative foot points of magenta (F2) and blue (F4) tubes are rooted in intruding the negative peninsula (i.e., N1) and N2 respectively, while the positive foot points of both tubes (F1 and F3) are rooted in P1.
Note that a part of the positive foot point of the blue tubes is rooted slightly closer to the intruding negative peninsula although still in P1; thus, we labeled the foot point as F'3.

\begin{figure}[ht!]
\epsscale{0.90}
\plotone{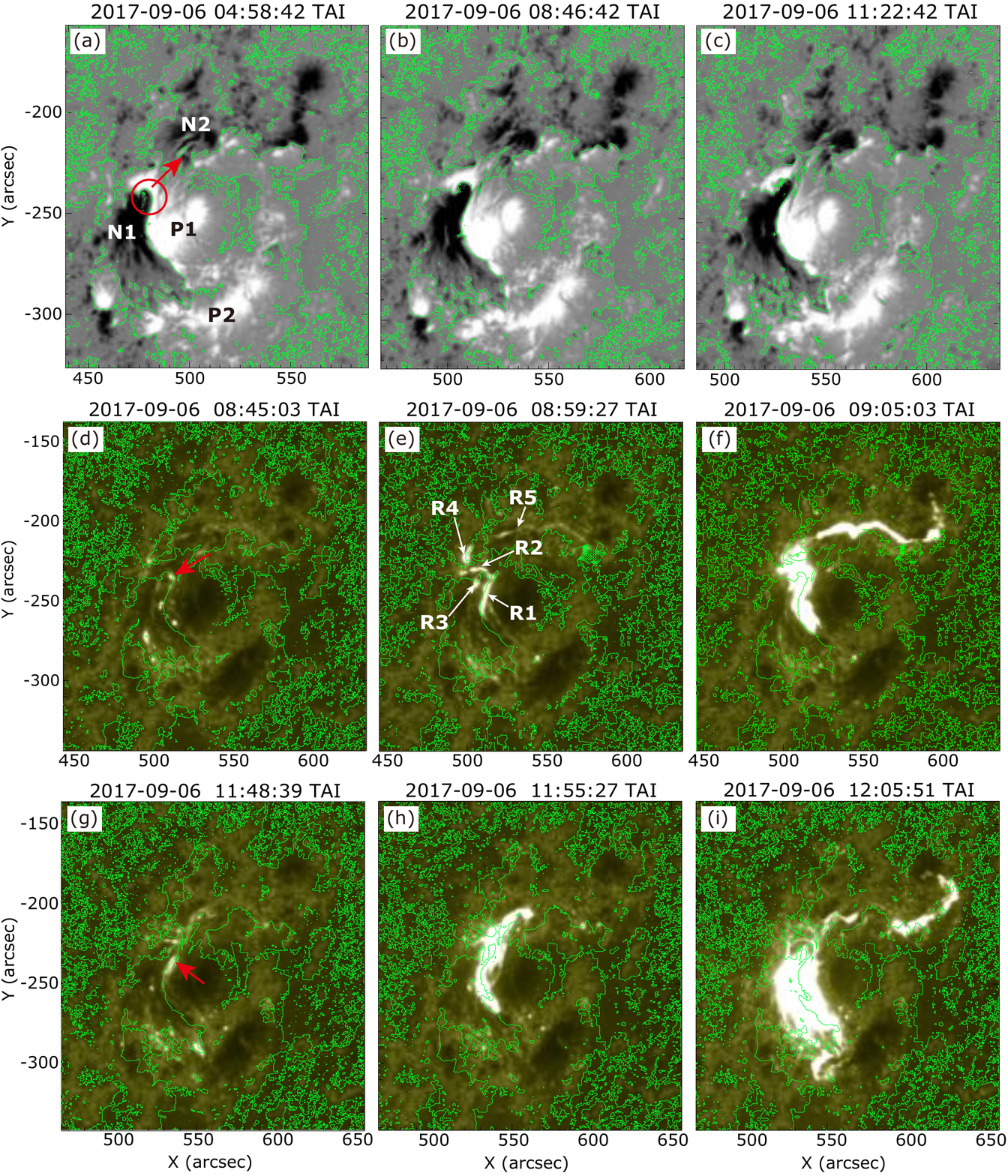}
\caption{
(a-c) Temporal evolution of the radial magnetic field $B_z$ observed by {\it SDO}/HMI SHARP.
The background white/black areas indicate the positive/negative polarities of $B_z$ in the range of ${\pm 1,000}G$, and the magnetic polarity inversion lines are indicated by the green lines.
The major polarities are labeled as N1 and N2 for the negative regions and P1 and P2 for the positive regions.
The intruding negative peninsula is denoted by the red circle in panel (a).
(d-i) Temporal variation of brightening in {\it SDO}/AIA 1600{\AA} in the range of 0-1,000 DN.
The left, middle, and right columns depict snapshots of the precursor brightening, initial flare ribbons, and enhanced flare ribbons, respectively.
Panels (d-f) and (g-i) correspond to the X2.2 and X9.3 flares, respectively.
The magnetic polarity inversion lines of $B_z$ are overplotted with the green lines.
The red arrows in panels (d, g) indicate the location of the last precursor brightening that is observed just before the X2.2 and X9.3 flare onsets, and the white arrows in panel (e) indicate the X2.2 initial flare ribbons.
The field-of-views of panels (d-i) are slightly larger than those of panels (a-c).
The supplemental animation of the AIA 1600 {\AA} (without the magnetic polarity inversion lines) is available.
}
\label{fig:AR_evolution}
\end{figure}

\begin{figure}[ht!]
\plotone{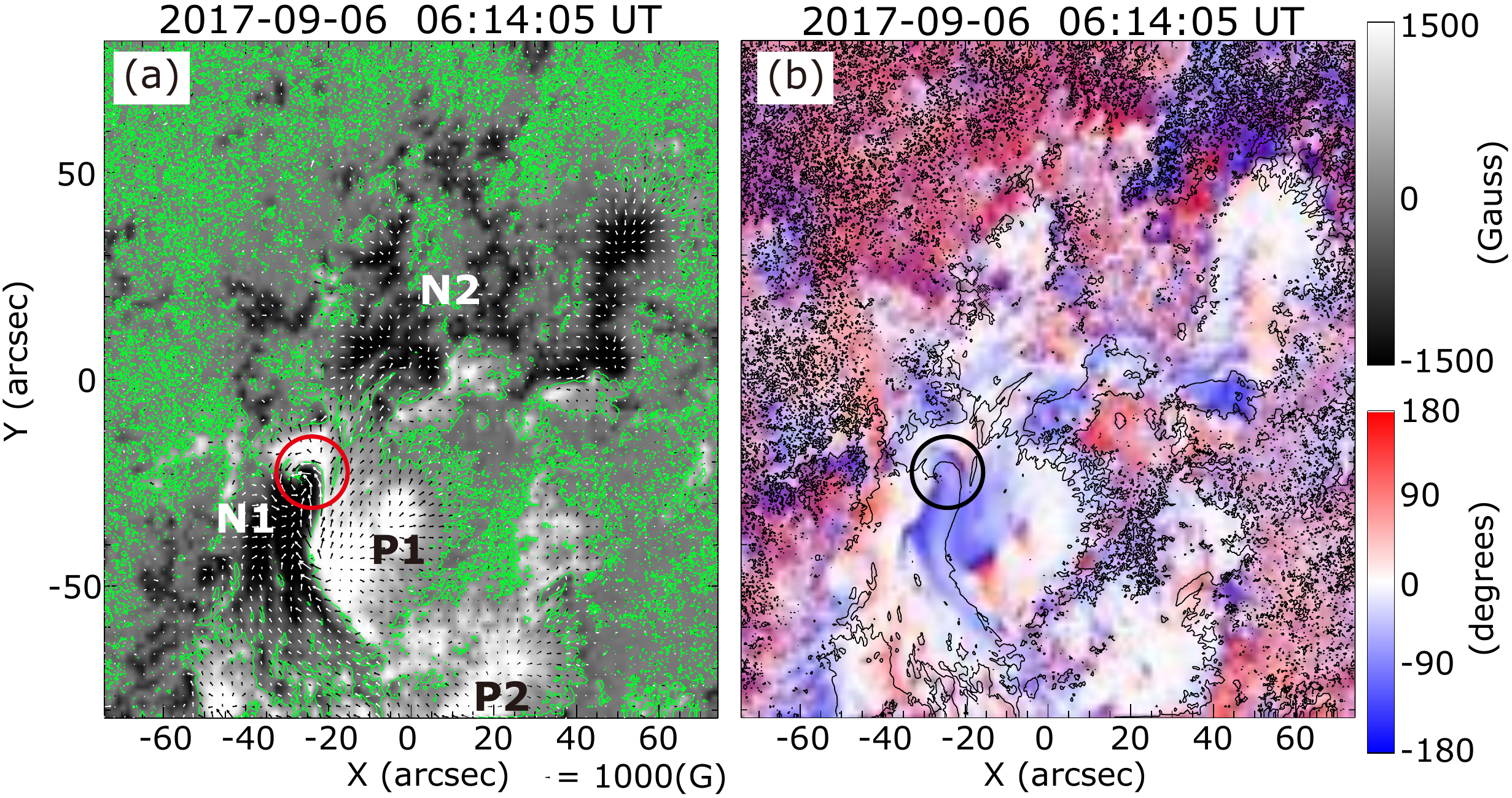}
\caption{
(a) Distribution of three components of the magnetic field ($B_x$, $B_y$, $B_z$) in AR 12673, which was observed by {\it Hinode}/SOT-SP fast-map scanning from 06:14:05 UT on September 6, 2017.
The background white/black and green lines are similar to those in Figure~\ref{fig:AR_evolution}(a-c), except that the background $B_z$ level is in the range of ${\pm 1,500}G$.
The small arrows indicate the orientation and strength of the horizontal components ($B_x$, $B_y$) of the photospheric magnetic fields, whose magnitudes are greater than 100G.
(b) Distribution of the magnetic shear angles relative to the potential field.
The red/blue regions indicates counter-clockwise/clockwise shear in the range of ${\pm 180^{\circ}}$.
The black lines represent the magnetic polarity inversion lines, similar to the green lines in panel (a).
The red/black circles in panel (a)/(b) denote the intruding negative peninsula.
Additionally, the southern half of the P2 is located outside of the SP scanning field-of-view.
}
\label{fig:SP_angles}
\end{figure}

\begin{figure}[ht!]
\plotone{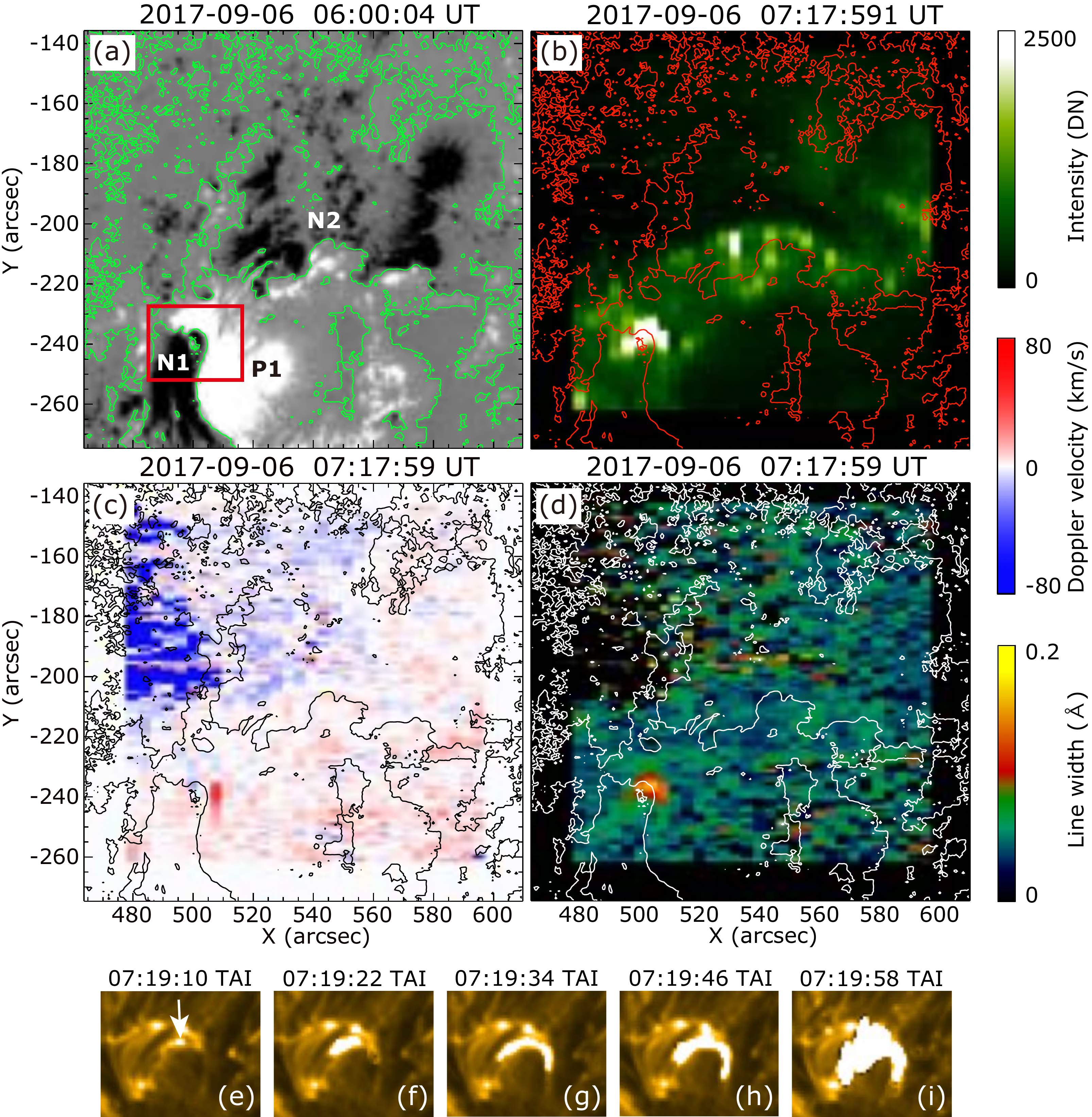}
\caption{
Intensity (upper-right, panel (b)), Doppler velocity (lower-left, panel (c)), and line width (lower-right, panel (d)) in \ion{Fe}{14} (264.7{\AA}) line observed by {\it Hinode}/EIS.
The start time of the scanning is 07:17:59 UT on September 6, 2017, and it takes approximately 3min to scan from the west to east (right to left) in the field-of-view.
The color scales for intensity, Doppler velocity, and line width are depicted on the right.
Panel (a) is a reference ${B_z}$ magnetogram observed by {\it SDO}/HMI SHARP at 06:00 UT on September 6, 2017, which is closest in time to the {\it Hinode}/EIS observation of panels (b-d).
The intensity, Doppler velocity, and line width maps and ${B_z}$ magnetogram are co-aligned with each other, and the green colored magnetic polarity inversion lines in panel (a) are overplotted on panels (b-d) denoted in red, black, and white lines, respectively.
Temporal evolution of the small cusp-shaped brightening in panel (b) is depicted in panels (e-i), which are coronal images captured by {\it SDO}/AIA 171{\AA} in the field-of-view of the red rectangle in panel (a).
The cusp-shaped brightening in panel (b) starts as a tiny brightening that is indicated by the white arrow in panel (e).
}
\label{fig:EIS_SHARP}
\end{figure}

The negative foot point F2 of the magenta tubes is pressed against the blue tubes as the intruding negative peninsula moves toward northwest.
This motion possibly causes ``push-mode'' magnetic reconnection \citep{yamada99} that forms a magnetic flux rope erupting with the X2.2 flare.
Figures~\ref{fig:schema}(a) and (b) illustrate the before and after versions of push-mode magnetic reconnection that varies magnetic topologies from F1-F2 /F3-F4 (the magenta/blue tubes) to F2-F3/F1-F4 (yellow tubes).
It is inferred that F1-F4 develops as a flux rope and the bright loop L1 in Figures~\ref{fig:AIA171fluxrope} (b) and (c) corresponds to overlying magnetic fields on the magnetic flux rope.
The overlying loop L1 becomes bright as the flux rope grows and rises toward its eruption, i.e., the X2.2 flare, owing to the intrusion of the intruding negative peninsula.
The initial flare ribbons then appear at each foot point as shown in Figure~\ref{fig:AR_evolution} (e).
The foot points F1 and F3 in Figure~\ref{fig:NLFFF} (b) correspond to the initial flare ribbons R1 and R2 on the positive polarity P1.
The other ribbons R3, R4, and R5 on the negative polarities N1 and N2 appear at the foot points of F2 and F4.
Here, the foot point F4 is rooted by both the flare ribbons R4 and R5.

Moreover, F'3-F4, which is a part of the pre-existing blue tubes and is denoted by the blue dashed line in Figure~\ref{fig:schema}(a), is also reconnected with the magenta tubes F1-F2.
F'3-F4 reconnects with F1-F2 earlier than F3-F4 because it is originally located closer to F1-F2 than F3-F4.
This magnetic reconnection with F'3-F4 and F1-F2 forms a small loop F2-F'3, which is denoted by the red dashed line in Figure~\ref{fig:schema}(b), and corresponds to the small cusp-shaped brightening that is observed in {\it Hinode}/EIS, XRT, and {\it SDO}/AIA 171 {\AA} (cf. Figures~\ref{fig:EIS_SHARP}, \ref{fig:XRT_MAG}, and \ref{fig:AIA171fluxrope}) before the X2.2 flare.

Figure~\ref{fig:NLFFF}(c) displays the magnetic field lines before the X9.3 flare.
The intrusion of the negative peninsula continues even after the X2.2 flare, and it further causes magnetic reconnection under the yellow tubes.
A part of yellow tubes then becomes unstable and is lifted.
Thus, the initial flare ribbons of the X9.3 flare appear under the yellow tubes as shown in Figure~\ref{fig:NLFFF} (d).
The yellow tubes further reconnect with the green tubes that are rooted in P2.
This magnetic reconnection causes variations in the magnetic topologies from F1-F4/F5-F6 (the yellow/green loops) to F4-F6/F1-F5 (thick and thin red loops), as illustrated in Figures~\ref{fig:schema}(c) and (d).
Furthermore, the initial flare ribbons propagate to the P2 region, where F6 is rooted in, and are elongated toward the south more than that of the X2.2 flare, as shown in Figure~\ref{fig:AR_evolution}(i).

The magnetic flux rope F1-F4, which is indicated by the thick yellow line in Figure~\ref{fig:schema}(c), might correspond to the dark tube-like structure that was observed in the {\it Hinode}/XRT and {\it SDO}/AIA 171 {\AA} images between the X2.2 and X9.3 flares.
The tube-like structure started to observed in {\it SDO}/AIA 171 {\AA} from 10:45 UT (approximately 1.5h before the X9.3 flare onset), while it was not seen in  {\it SDO}/AIA 171 {\AA} and {\it Hinode}/XRT before and just after the X2.2 flare.
Furthermore, we observed precursor brightening that is elongated along the magnetic polarity inversion line between P1 and N1 before the X9.3 flare onset, as denoted by the red arrow in Figures~\ref{fig:AR_evolution}(g) and \ref{fig:AIA171fluxrope}(e).
Those observational results suggest that successive magnetic reconnection under the flux rope F1-F4 occur and reinforce the flux rope \citep[cf.][]{fan10}.

Therefore, according to the results obtained from observational data analysis, the intrusion of the negative peninsula that presses the magenta magnetic field lines to the blue magnetic field lines promote push-mode magnetic reconnection that causes the X2.2 flare.
Moreover, the successive intrusion of the negative peninsula further plays an important role to cause successive magnetic reconnection and to hasten the eruption of the yellow magnetic flux rope by an MHD instability.
\citet{zou20} also stated the importance of the first X2.2 flare reconnection to facilitate the rise of the flux rope and its eruption, i.e., the X9.3 flare, by torus instability.
The magnetic polarity inversion line between P1 and N1 has strong magnetic gradient, resulting in strong electric current.
Therefore, there is a possibility that magnetic reconnection between sheared magnetic fields over the P1-N1 magnetic polarity inversion lines occurs and forms a magnetic flux rope by consuming a considerable amount of time, even without the intrusion of the negative peninsula.
Nevertheless, our results suggest that the intrusion of the negative peninsula encourages the formation and destabilization of magnetic flux ropes both in the X2.2 and X9.3 flares.

\subsection{Interpretation of the Local and Transient Downflow} \label{sec:rdiscussion_downflow}

\begin{figure}[ht!]
\plotone{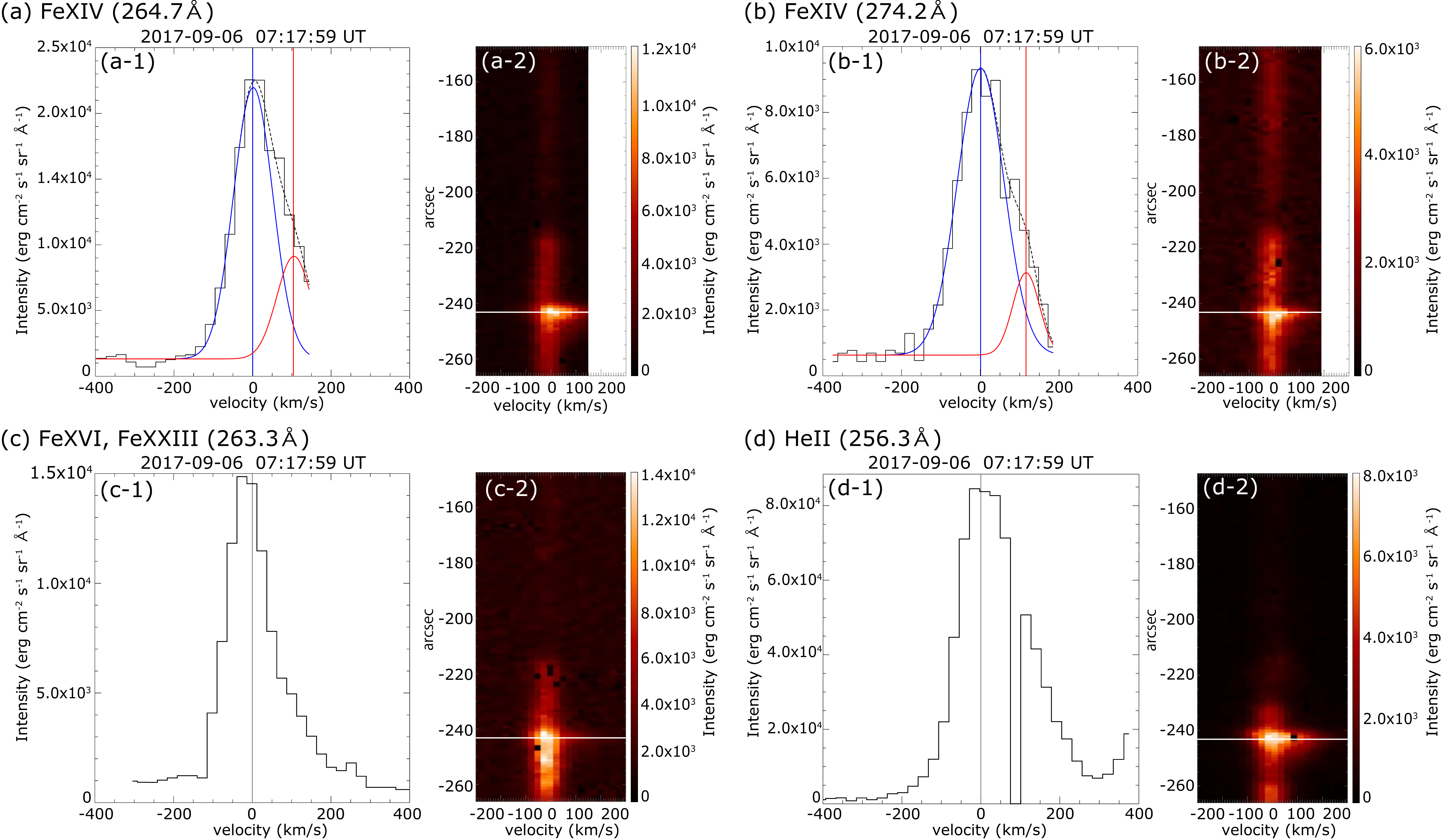}
\caption{
Spectral line profiles and spectral images at the single site of the red-shifted signal in Figure~\ref{fig:EIS_SHARP}(c).
Each panel is corresponds to the \ion{Fe}{14} (264.7 {\AA}), \ion{Fe}{14} (274.2 {\AA}), \ion{Fe}{16}/\ion{Fe}{23} (263.3 {\AA}), and \ion{He}{2} (256.3 {\AA}) lines, respectively.
The intensity profiles along the horizontal white line-cut in the spectral images on the right are plotted in the images on the left.
The vertical axes of the profiles represent intensity, and the vertical axes of the spectral images are represented in spatial arcseconds along the slit.
The horizontal axes represent the velocities resulting from single (in panels (c) and (d)) and double (in panels (a) and (b)) Gaussian fitting.
The intensity range of the spectral images is depicted in the color-scale on the right of each panel.
The histogram with the black solid line depicts the intensity of each spectral line.
The black broken lines are the result of Gaussian fitting, and the blue/red curves denote the first/second components obtained from double Gaussian fitting for the intensity profiles in panels (a-1) and (b-1).
The Doppler velocities of the second components relative to the first components are approximately $103 \pm 8 km/s$ for 264.7 {\AA} and $115 \pm 16 km/s$ for 274.2 {\AA}, respectively.
}
\label{fig:EISprofile}
\end{figure}

\begin{figure}[ht!]
\plotone{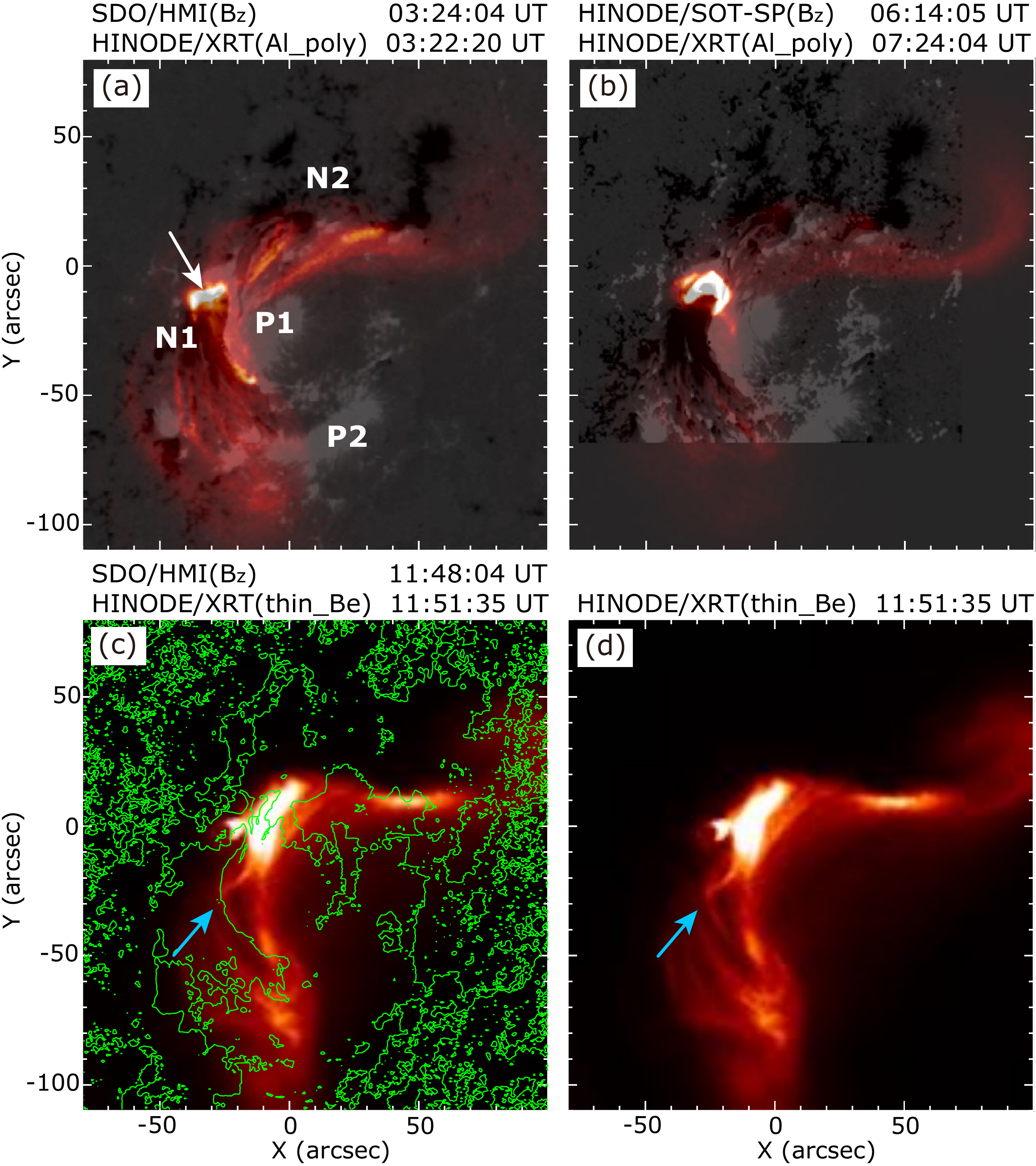}
\caption{
Temporal variation of high-temperature coronal structures observed by {\it Hinode}/XRT.
The red-colored X-ray images are displayed over the radial magnetic field ($B_z$) images that are closest in time to the X-ray images in panels (a) and (b).
The radial magnetic field image is obtained from the {\it SDO}/HMI SHARP series in panel (a), whereas the radial field image in panels (b) is obtained by {\it Hinode}/SOT-SP, because of the data gap in the {\it SDO} observation.
The white arrow in panel (a) indicates a small bright loop in the north of the intruding negative peninsula.
The background {\it Hinode}/XRT images in panels (c) and (d) are the same, whereas the magnetic polarity inversion lines of the radial magnetic field ($B_z$ obtained from {\it SDO}/HMI SHARP) are overlaid only in panel (c), depicted as green lines.
The sky-ble arrows in panels (c) and (d) indicate a dark tube-like structure along the magnetic polarity inversion line between P1 and N1.
}
\label{fig:XRT_MAG}
\end{figure}

\begin{figure}[ht!]
\centering
\includegraphics[width=11cm,clip]{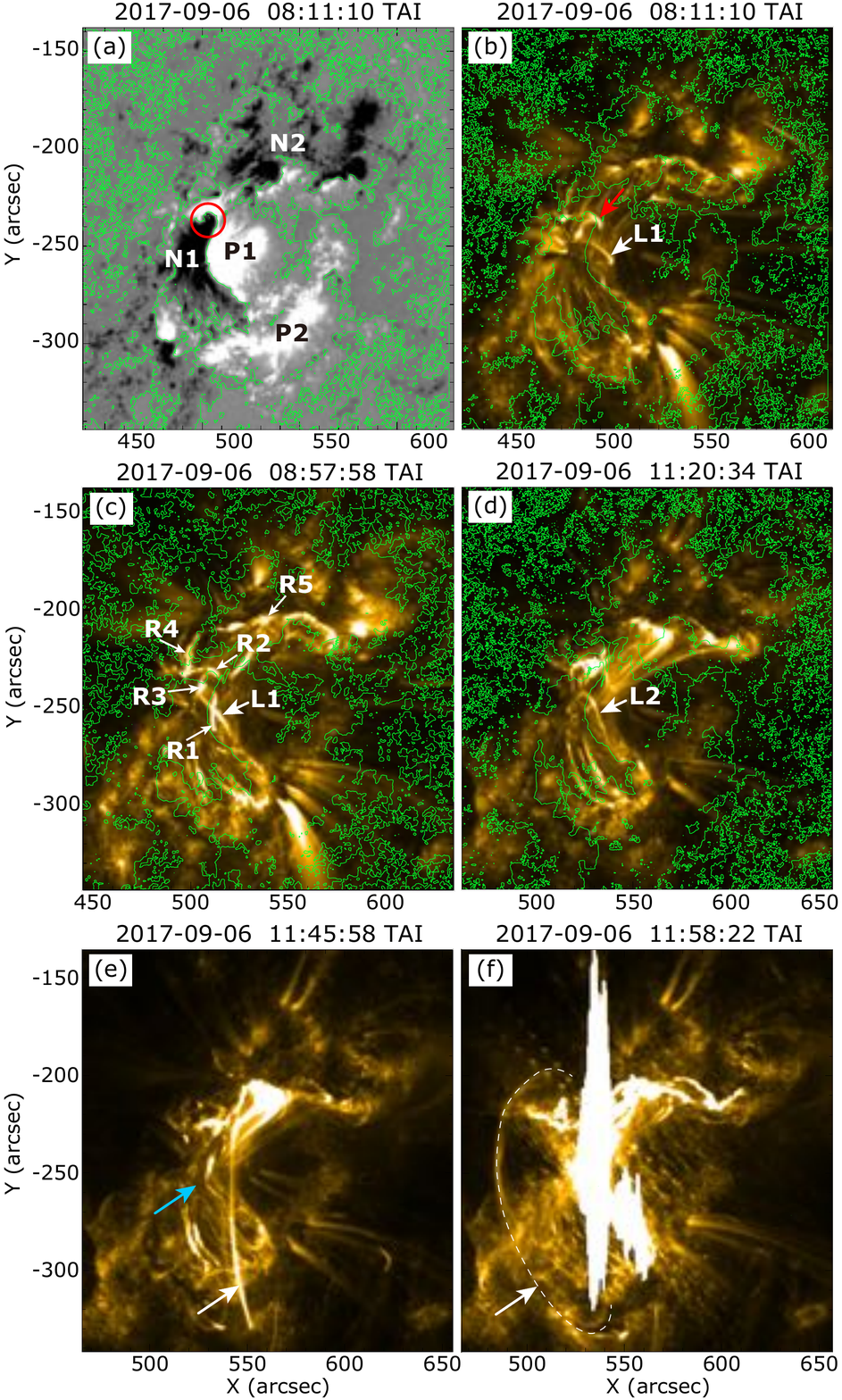}
\caption{
Temporal variation of the {\it SDO}/AIA 171 {\AA} images, whose characteristic emission temperature is $log(T)\sim5.8$.
Panel (a) depicts a reference magnetic field image at 08:11 UT on September 6, 2017, the same time to panel (b).
The magnetic polarity inversion lines of the radial magnetic field ($B_z$) obtained from {\it SDO}/HMI SHARP that are closest in time to the background {\it SDO}/AIA 171 {\AA} images are overlaid in panels (b-d).
The intruding negative peninsula is indicated by the red circle in panel (a).
The red arrow in panel (b) indicates the cusp-shaped bright loop that are seen in Figures~\ref{fig:EIS_SHARP}(b) and \ref{fig:XRT_MAG}(b).
The white arrows in panels (b) and (c) are indicate the same bright loop L1 at different times before the X2.2 flare onset.
Another white arrows in panel (c) indicates the initial flare ribbons of the X2.2 flare that is seen in Figure~\ref{fig:AR_evolution}(e).
The white arrow in panel (d) points to the bright loop L2 that appears between the X2.2 and X9.3 flares.
In panel (e), a dark tube-like structure along the magnetic polarity inversion line between P1 and N1, which is seen in Figure~\ref{fig:XRT_MAG}(c, d), is indicated by the sky-blue arrow.
The white arrows in panels (e) and (f) indicate the same long north-south loop before and after the X9.3 flare onset, and the white dashed line in panel (f) outlines the expanded north-south loop.
The supplemental animation of the {\it SDO}/AIA 171 {\AA} (without the magnetic polarity inversion lines) is available.
}
\label{fig:AIA171fluxrope}
\end{figure}

\begin{figure}[ht!]
\plotone{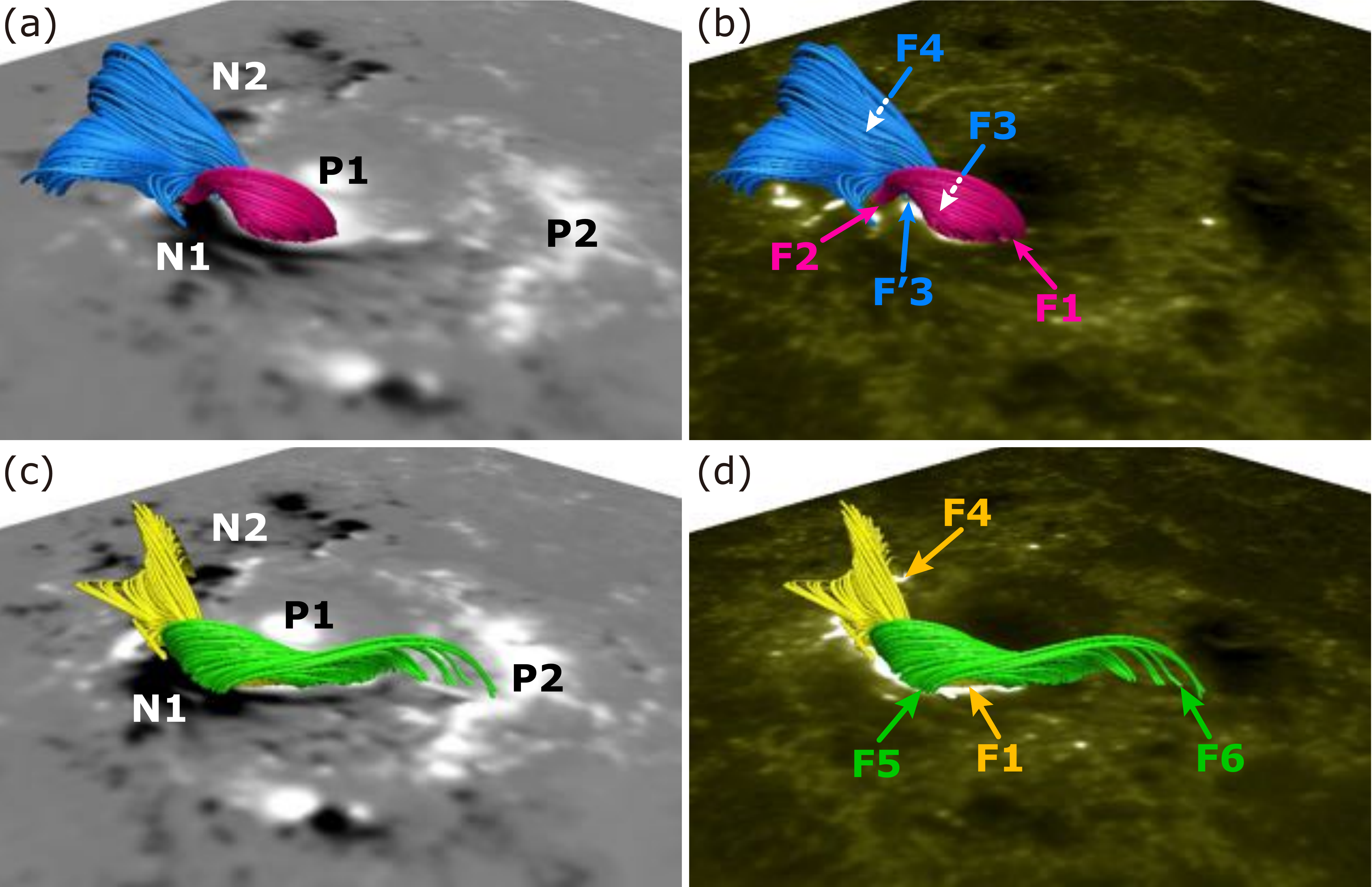}
\caption{
Three-dimensional coronal magnetic field structures before the onset of the X2.2 and X 9.3 flares, which are reconstructed by the NLFFF extrapolation.
All the images depict AR as seen from the southeast (i.e., from the bottom-left in the images of Figure~\ref{fig:AR_evolution}).
The upper two panels (a) and (b) are correspond to the magnetic field structure before onset of the X2.2 flare (calculated from {\it SDO}/HMI SHARP on 08:36 UT on September 6, 2017), whereas the bottom two panels (c) and (d) are correspond to that before the onset of the X9.3 flare (calculated from {\it SDO}/HMI SHARP on 11:00 UT on September 6, 2017).
The different colored tubes indicate magnetic field lines with different magnetic connectivities.
Panels (a) and (c) depict the side view of the magnetic fields with a bottom image of the radial magnetic field ($B_z$ obtained from {\it SDO}/HMI SHARP).
Panels (b) and (d) depict the same view with the background of AIA 1600 {\AA} images at the time of the initial flare ribbons at 08:59:27 and 11:55:27 UT on September 6, 2017,  which are the same images as Figures~\ref{fig:AR_evolution} (e) and (h), respectively.
}
\label{fig:NLFFF}
\end{figure}

\begin{figure}[ht!]
\plotone{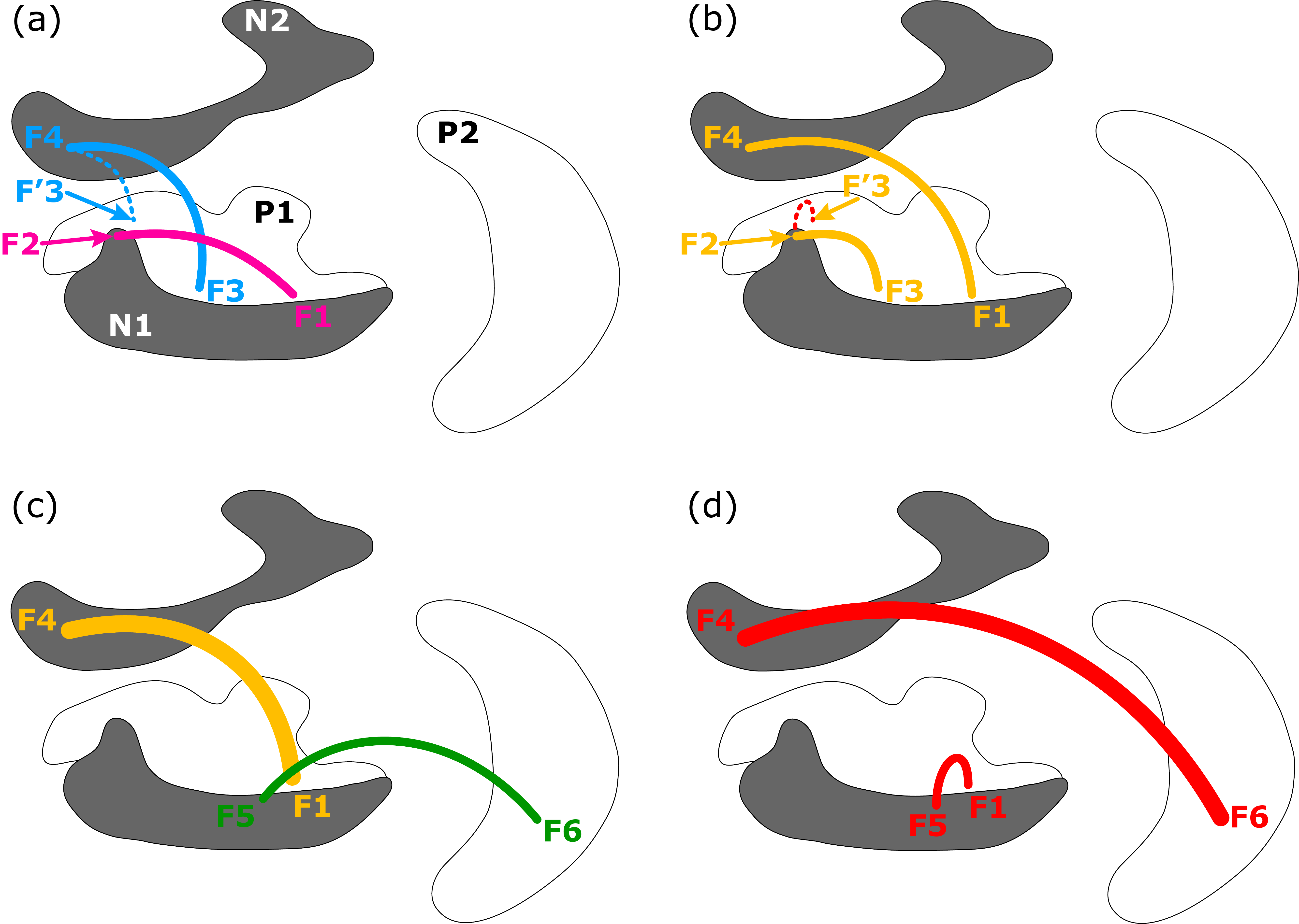}
\caption{
Schematics of magnetic field lines before and after the onset of the X2.2 and X9.3 flares.
All the images depict AR as seen from the southeast (i.e., from the bottom-left in the images of Figure~\ref{fig:AR_evolution}), similar to that in Figure~\ref{fig:NLFFF}.
The colored lines in panels (a) and (c) represent the same colored tubes in Figure~\ref{fig:NLFFF}(a) and (c), and are correspond to the snapshots before the onset of the X2.2 and X9.3 flares, respectively.
The red and yellow solid lines in panels (b) and (d) indicate the magnetic field lines after the onset of the X2.2 and X9.3 flares.
The small red dashed line loop corresponds to the cusp-shaped brightening depicted in Figures~\ref{fig:EIS_SHARP} and \ref{fig:XRT_MAG}.
}
\label{fig:schema}
\end{figure}

In this study, we found local and transient downflow related to the cusp-shaped brightening as we described in Section~\ref{sec:results_local}.
Generally, Lorentz force driven flows such as reconnection out flow do not exhibit clear temperature dependence while pressure gradient driven flows such as chromospheric evaporation exhibit the strong temperature dependence \citep[e.g.,][]{fisher85, watanabe10, imada15}.
We observed the downflow of $\sim 100 km/s$ at both high temperature Fe lines ($log(T)\sim6.3-6.4$) and low temperature \ion{He}{2} line ($log(T)\sim4.9$).
The temperature is believed to rises proportionately to height in the solar atmosphere in the region where is higher than the chromosphere \citep[e.g.,][]{vernazza81}.
Therefore, each ion's sensitive temperature is suggested to correspond to the formation height of each emission line.
Further, the downflow illustrate in Figure~\ref{fig:EIS_SHARP}(c) is driven by Lorentz force and that there is an energy input, such as magnetic reconnection, that occurred in a higher atmosphere than formation height of the {\it Hinode}/EIS Fe lines shown in Figure~\ref{fig:EISprofile}.
The velocity of the downflow was approximately $100 km/s$, which is significantly lower than the Alfv$\acute{e}$n velocity (order of $1,000 km/s$) in the corona, in all the Fe and He lines shown in Figure~\ref{fig:EISprofile}.
The small cusp-shaped high-temperature loop that was observed using {\it Hinode}/XRT in Figure~\ref{fig:XRT_MAG}(b) and magnetic topology in Figure~\ref{fig:NLFFF} also supports this interpretation.

As we discussed in Section~\ref{sec:rdiscussion_scenario}, the push-mode magnetic reconnection between the magenta and blue sheared magnetic fields due to the intrusion of the negative peninsula can form a small magnetic loop connecting P1 and N1 at the tip of the intruding negative peninsula, as well as the cusp-shaped brightening.
Moreover, downflow was observed in the west leg of the cusp rather than the top of the cusp, as seen in Figure~\ref{fig:EIS_SHARP}(b, c).
Therefore, it is suggested that the downflow indicates a secondary flow in which coronal plasma decelerate and fall down along the coronal loop \citep[e.g., ][]{imada13, polito18}, rather than the reconnection out-flow itself \citep[e.g., ][]{yokoyama98}.

\citet{bamba17b} found similar local and transient upflow using lower temperature lines ($log(T)\sim4.0-4.8$), which were observed by the {\it IRIS}\footnote{{\it Interface Region Imaging Spectrograph}} satellite \citep{depontieu14}, than the {\it Hinode}/EIS observed lines in this study.
In their case, temperature independent strong upflow with a velocity of up to $100 km/s$ was observed, corresponding to a brightening in the chromosphere before an X-class flare.
They interpreted that the local and transient upflow suggest the presence of an energy input from magnetic reconnection in a lower temperature region lower than the middle chromosphere.
Unfortunately, data from IRIS, which can observe emission lines from a lower temperature atmosphere such as that used in \citet{bamba17b}, was not available for this AR 12673.
Even so, it is suggested that our downflow is driven by Lorentz force because both the high-temperature Fe lines and low-temperature \ion{He}{2} line showed red-shifted signals with the velocities of approximately $100 km/s$.

Moreover, in \citet{bamba17b}, the local and transient upflow, which was observed approximately 30min before an X-class flare onset by {\it IRIS}, is reported as a proxy of precursory phenomenon that directly represents the trigger of an X-class flare.
Their upflow was very transient but the corresponding brightening in the chromosphere continued until 10min before an X-class flare onset.
In our case, it seems that the location of the cusp-shaped brightening and downflow corresponds to the location of the push-mode magnetic reconnection, which plays a crucial role in triggering the X2.2 flare.
However, it is unreasonable to assume that those features represent the direct trigger of the X2.2 flare because those were observed more than 1.5h before the flare onset.
Therefore, it is suggested that the cusp-shaped brightening and downflow are proxies of the push-mode magnetic reconnection that gradually occurs at the tip of the intruding negative peninsula due to its intrusion from before the X2.2 flare.
As we described in Section~\ref{sec:rdiscussion_scenario}, the magnetic connectivity of the cusp-shaped brightening seems to be formed by push-mode magnetic reconnection between the magenta and blue magnetic arcades in Figure~\ref{fig:NLFFF} (a) and (b).
This topological change possibly forms a magnetic flux rope corresponding to the F1-F4 loop and is erupted with the X2.2 flare.

\section{Summary} \label{sec:summary}

In the present study, we examined the onset process of the consecutive X2.2 and X9.3 flares because the X9.3 flare is the largest flare in the solar cycle 24 and impacted the space weather of the Earth.
We analyzed observational data from the {\it Hinode} and {\it SDO} satellites and interpreted the results using supplemental NLFFF extrapolation.
As a result, we found that the key feature to trigger the consecutive X-class flares is the intrusion of the negative peninsula towards the neighboring opposite polarity region in the northwest.

The intrusion is crucial to promote the ``push-mode'' magnetic reconnection \citep{yamada99} between the sheared magnetic field along the magnetic polarity inversion line between the major bipoles P1 and N1 in the AR, and to form a twisted magnetic flux rope.
The ``push-mode'' magnetic reconnection is introduced based on reconnection experiments performed in laboratories, as well as another ``pull-mode'' magnetic reconnection.
Both the push-mode and pull-mode reconnection can occur on a vertical electric current sheet formed between large-scale sheared magnetic fields by converging or diverging flows, respectively.
Flux cancellation \citep{van89, green11} on the photosphere corresponds to the push-mode reconnection. 

Therefore, we propose an onset scenario of the consecutive X-class flares as follows: the flux rope that is formed and evolved by the intrusion of the negative peninsula erupts and causes the X2.2 flare; the successive intrusion of the negative peninsula promotes further breaking of the magnetic equilibrium, leading to reinforcement of the magnetic flux rope; the flux rope then erupts due to destabilization of the magnetic field and causes the X9.3 flare.

We presented observational evidences that support the above scenario.
First, magnetic field data showed the intrusion of a part of a major negative sunspot (i.e., negative peninsula) towards the opposite polarity region in the northwest, and precursor brightening was correspondingly observed at the tip of the intruding negative peninsula.
Second, the cusp-shaped brightening and corresponding downflow that can be proxies of push-mode magnetic reconnection in the presented magnetic topology during the intrusion of the negative peninsula were detected by both imaging and spectroscopic observations.
Third, the magnetic flux rope was observed in extreme-ultraviolet and X-rays between the X2.2 and X9.3 flares, and it erupted as the X9.3 flare.
However, it was difficult to clearly identify whether the flux rope formed by X2.2 flare was necessary for the X9.3 flare occurrence.
We require further studies to examine reveal this problem.

We further demonstrated the usefulness of complementary observational data analysis by combining observations, such as {\it Hinode} and {\it SDO} data.
Those satellites are unique owing to their high spatial resolution, high temporal cadence, and multiple observational methods such as imaging, spectroscopy, and spectropolarimetry.
It would be useful to investigate the detailed physical mechanisms of flare triggering through multiple observations and MHD simulations.
Therefore, further understanding of the physical mechanism of flare triggers with a combination of observational and numerical studies is expected.


\acknowledgments

The authors appreciate all the people working on {\it Hinode} for the perfect operation to obtain such high-quality data of AR 12673.
Moreover, the authors thank to the proposers of the observations that were postponed owing to the AR 12673 flare observation for their understanding.
The HMI and AIA data have been used courtesy of NASA/{\it SDO} and the AIA and HMI science teams.
The authors are grateful for the science teams.
This work was partly carried out at the Hinode Science Center at NAOJ and Nagoya University, Japan, and the Solar Data Analysis System (SDAS) operated by the Astronomy Data Center in cooperation with the Hinode Science Center of the NAOJ.
A part of this study was carried by using the computational resource of the Center for Integrated Data Science, Institute for Space-Earth Environmental Research, Nagoya University through the joint research program.
The authors and part of this work were supported by MEXT/JSPS KAKENHI Grant Numbers JP15H05812, JP15H05814, JP15H05816, and JP15K21709.
YB deeply appreciate the many instructive comments from Dr. D. H. Brooks and Dr. K. D. Leka, discussion with Dr. K. Kusano and Dr. J. Muhamad, and the kind technical support from Dr. T. Hara.
The authors thank the anonymous referee for their valuable comments that improved the clarity of the manuscript.
We would like to thank Editage (www.editage.com) for English language editing.


\vspace{5mm}
\facilities{Hinode (SOT, EIS, and XRT), SDO (HMI and AIA)}
\software{SSW, SPEDAS, VAPOR}

\appendix \label{sec:appendix} 

\begin{figure}[ht!]
\plotone{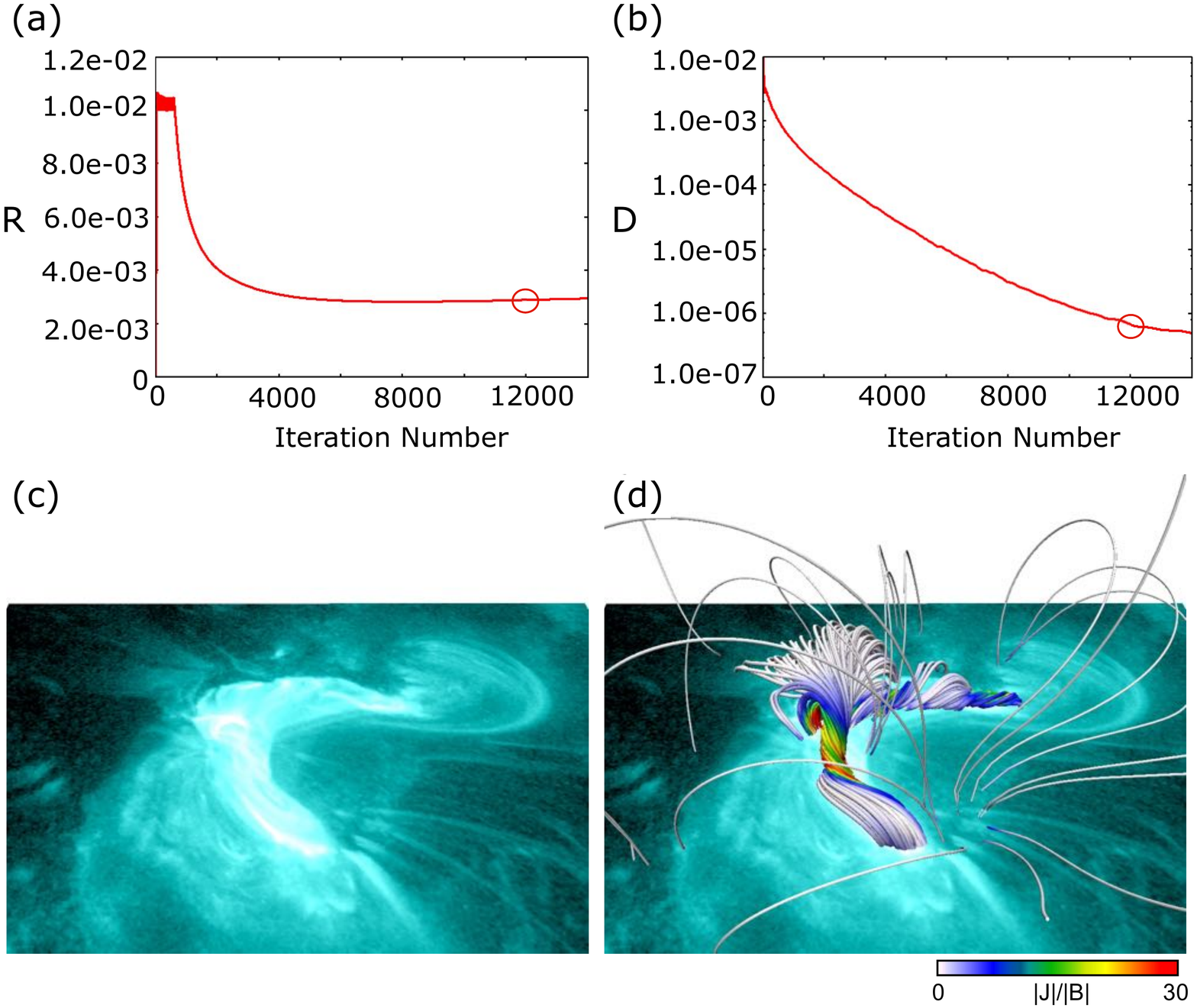}
\caption{
(a)-(b) Temporal evolution of residual Lorentz-force and divergence of magnetic field, which are defined as $R=\int |\vec{J}\times\vec{B|}^2dV$ and $D=\int |\vec{\nabla}\cdot\vec{B}|^2$dV, respectively, where $dV$ is a volume element.
We considered the magnetic field as the NLFFF at $12,000$ iterations denoted by the red circle.
(c) EUV image obtained from AIA 131 {\AA} observed at 08:36 UT on September 6, 2017.
(d) The NLFFF extrapolated from the photospheric magnetic field observed at 08:30 UT on September 6, 2017, is superimposed on the image depicted in (c).
}
\label{fig:appendix}
\end{figure}

We evaluate the NLFFF model extrapolated from the photospheric magnetic field, focusing on data taken at 02:30 UT on September 6, 2017.
The temporal evolution of residual Lorentz-force $R = \int |\vec{J}\times\vec{B}|^2dV$ and divergence of magnetic field $D =\int |\vec{\nabla}\cdot\vec{B}|^2dV$ during the NLFFF calculation are plotted in Figures~\ref{fig:appendix}(a) and (b), where $dV$ is a volume element.
Both profiles decrease as the evolution progresses.
We change the boundary magnetic field during the calculation; in particular, the horizontal component is according to a linear combination, $\vec{B}_{t} = \gamma \vec{B}_{t(obs)} + (1-\gamma)\vec{B}_{t(pot)}$, where $\vec{B}_{t(obs)}$ and $\vec{B}_{t(pot)}$ correspond to the horizontal component of the observed and potential magnetic fields measured at the bottom surface, respectively.
$\gamma$ increases as per the following formulation: $\gamma = \gamma + \delta \gamma$ falls below $R_{min}$, where $R_{min}$ is set at $1.0\times 10^{-2}$ as a given parameter.
If $\gamma$ satisfies $1.0$, $\vec{B}_{t}$ is consistent with observed magnetic field $\vec{B_{t(obs)}}$.
As shown in Figure~\ref{fig:appendix}(a), the value of $R$ saturates around $1.0 \times 10^{-2}$ at a very early stage of the calculation because the horizontal magnetic field at the bottom boundary is changing from a potential field to the observed magnetic field.
When $\gamma$ is $1$, the magnetic field is relaxed towards to the NLFFF.
Since the value of $R$ slightly increases after $8,000$ iterations, we took the magnetic field at $12,000$ iterations, where the value of $R$ falls below $1.0 \times 10^{-6}$.
Figures~\ref{fig:appendix}(c) and (d) show the extreme ultra violet (EUV) image taken by AIA at 131{\AA} and the NLFFFF superimposed on the image, respectively.
The colors of the field lines correspond to the value of $|\vec{J}|/|\vec{B}|$, and the strong brightening region shown in the EUV image is well captured by the NLFFF.





\end{document}